\documentclass{aa}
\usepackage{graphics}
\usepackage{epsfig}
\begin{document}
\thesaurus{08(09.02.1; 09.03.1; 10.15.1; 10.19.1)}
\title{High-latitude molecular clouds and near-by OB associations}
\author{H. C. Bhatt\thanks{E-mail:hcbhatt@iiap.ernet.in}}
\institute{Indian Institute of Astrophysics, Bangalore 560 034, India}
\offprints{H. C. Bhatt}
\date{Received / Accepted }
\maketitle
\begin{abstract}
The Galactic distribution of the high-latitude molecular clouds is considered.
It is suggested that the majority of these clouds are clustered in two large 
shells around the two closest OB associations, Per OB3/Cas-Tau and Sco OB2. 
The most prominent shell of high-latitude clouds is centred around the Per OB3/
Cas-Tau association which is also at the centre of the Gould's Belt. However, 
the Per OB3/Cas-Tau group of high-latitude clouds appears as an elliptical 
shell at nearly right angles to the plane of the Gould's Belt. Its kinematic 
age ($\sim$ 10 {\rm Myr}) is much smaller than the expansion age ($\sim$ 35 
{\rm Myr}) of the ring of intertellar matter associated with the Gould's Belt.
It is suggested that while, as is generally understood, the larger expanding 
ring of gas associated with the Gould's Belt was created by stellar winds and 
supernova explosions of the more massive OB stars of the Per OB3/Cas-Tau 
association $\sim$ 35 {\rm Myr} ago, the Per OB3 shell of high-latitude clouds 
was formed from the back-falling gas as it was swept up by a more recent 
supernova explosion of an early B type star in Per OB3 $\sim$ 10 Myr ago. More 
distant OB associations also produce similar shells of clouds at large heights 
from the Galactic plane having relatively smaller angular sizes due to their 
greater distances. These can be seen as higher latitude extensions of cloud 
complexes at lower Galactic latitudes as, for example, in the case of Vela OB2.


\end{abstract}

{\bf 1. Introduction}\\

Molecular clouds are an important constituent of the interstellar medium 
(ISM) of our Galaxy. They are found to be generally concentrated along
the Galactic equator (Lynds 1962; Feitzinger \& Stuwe 1984; Hartley et
al. 1986). With a relatively small scale height of $\sim$ 75 {\rm pc}
(Sanders et al. 1984; Blitz 1990) perpendicular to the Galactic plane and 
typical distances within $ \sim$ 1 {\rm kpc} these clouds are to be found 
at Galactic latitudes within $\sim 10^{\degr}$. In optical surveys (for 
obscuring dark clouds) only a handful of clouds could be seen at Galactic 
latitudes $|b| \geq 25^{\degr}$. However, in recent years a number of molecualr
clouds at such high Galactic latitudes have been detected, mostly in CO surveys
(Blitz et al. 1984; Magnani et al. 1985, hereafter MBM; Keto \& Myers 1986; 
Magnani et al. 1996; Hartmann et al. 1998; Magnani et al. 2000 and references
therein).\\

The high-latitude molecular clouds have CO column densities $\geq 10^{15}$
{\rm cm$^{-2}$} and are readily detected in the CO ($J = 1-0$) transition.
However, over most of cloud surface, these clouds are characterized by low
visual extinctions, making it difficult to identify them on the Palomar
Observatory Sky Survey (POSS) plates and similar other optical surveys.
Most of the high-latitude clouds are therefore classified as translucent
clouds (van Dishoeck et al. 1991). From a literature search, Magnani et al.
(1996, hereafter MHS) produced a catalogue of the high-latitude molecular 
clouds (with galactic latitude $|b| \geq 25^{\degr}$) that contains 120 of 
these objects. To search for molecular clouds at $|b| \geq 30^{\degr}$ more 
complete surveys in CO have recently been conducted in the northern Galactic 
hemisphere (NGH) by Hartmann et al. (1998, hereafter HMT), and in the southern 
Galactic hemisphere (SGH) by Magnani et al. (2000, hereafter MHHST). The NGH 
survey by HMT detected CO emission along 26 lines of sight, mostly 
corresponding to high-latitude clouds already known, and yielded only 2 new 
molecular clouds. The survey by MHHST found 144 distinct CO emission lines 
along 133 lines of sight in the SGH. Of these, 58 are new and 75 are associated
with 26 previously catalogued high-latitude molecular clouds situated within 
the survey boundaries. MHS derived a velocity dispersion 5.8 {\rm km s$^{-1}$} 
for the clouds in their catalogue if seven intermediate velocity objects are
excluded and 9.9 {\rm km s$^{-1}$} otherwise. The scale height and the mean
distance of the clouds implied by these velocity dispersions are 124 
{\rm pc} and 150 {\rm pc}, and 210 {\rm pc} and 260 {\rm pc} respectively.
For the clouds detected in the recent NGH and SGH surveys MHHST derive a
velocity dispersion $\sim $ 7 {\rm km  s$^{-1}$} leading to the values of
$\sim $100 {\rm pc} and $\sim $140 {\rm pc} for the scale height and mean
distance respectively. Thus the high-latitude clouds are relatively local 
clouds. Individual clouds range in size from less than $\sim 10^{-1}$ {\rm pc} 
to $\sim 10^{1}$ {\rm pc} and in mass from $\sim 10^{-1}$ $M_{\sun}$ to 
$\sim 10^{3}$ $M_{\sun}$, and may contribute $\sim 10\%$ to $\sim 20\%$ by 
mass to the molecular gas content in the local ISM.\\
 
The high-latitude molecular clouds are mostly gravitationally unbound (MBM) 
and could be in pressure equilibrium with the ISM (Keto \& Myers 1986). 
Distance estimates for these clouds ($\sim 10^{2}$ {\rm pc}) are similar to 
the mid-plane distances to the inner edges of the rarefied region in the local
ISM called the Local Bubble (e.g., Welsh et al. 1994; Breitschwerdt et al. 
1996) that is thought to be produced either by supernova explosions or stellar 
winds from OB associations. It has also been suggested (e.g., Elmegreen 1988) 
that the high-latitude clouds could be formed as condensations in swept-up
HI shells. The structure of the Local Bubble at high Galactic latitudes is 
still uncertain. If the high-latitude clouds are indeed associated with the 
Local Bubble, then a study of the spatial distribution of the high-latitude 
molecular clouds can also help delineate the boundaries of the Local Bubble.\\

In this paper we consider the spatial distribution of the high-latitude
molecular clouds. In Sect. 2 it is shown that a majority of all the known
high-latitude clouds can be assigned to two large shells centred around
the two nearest and youngest OB associations, namely the Per OB3/Cas-Tau 
and the Scorpius-Centaurus (Sco OB2) associations. Sect. 3 discusses
the kinematics of the Per OB3/Cas-Tau shell of high-latitude
clouds, its expansion and relationship with the Gould's Belt. Similar
distributions of clouds extending to large heights from the Galactic plane
around other OB associations are also shown. Sect.4 summarizes our
conclusions.\\

{\bf 2. The galactic distribution of high-latitude molecular clouds}\\

The MHS catalogue lists 120 high-latitude molecular clouds at Galactic 
latitudes $|b| \geq 25^{\degr}$. At Galactic latitudes $|b| \geq
30^{\degr}$ HMT list 26 detections of CO emission from the NGH survey,
while MHHST list detections along 133 lines of sight in the SGH. Some of
these detections correspond to clouds in the MHS catalogue, while the
others are new. Some MHS clouds in the regions of the NGH and SGH
surveys, especially the smaller ones, were missed by the NGH and SGH
surveys (MHHST). A few more optically identified clouds with $|b|$ $\geq
\sim 25^{\degr}$, that are not listed in the MHS, HMT and MHHST catalogues, 
can be found in Lynds (1962), Hartley et al. (1986) and Clemens \& Barvainis 
(1989). Also, the updated version of the Lynds (1962) catalogue, available at 
CDS, Strasbourg, lists one new cloud at galactic longitude $l = 147.5^{\degr}$ 
and latitude $b = + 50.1^{\degr}$. These additional high-latitude clouds from 
the updated Lynds' catalogue, Hartley  et al. (1986) and Clemens \& Barvainis 
(1989) presently lack molecular CO observations, but are likely to be molecular
clouds similar to the other high-latitude molecular clouds. These additional 
high-latitude clouds are listed in Table 1 which gives their designations 
(L for Lynds clouds, DCld for dark clouds from Hartley et al. (1986), CB for 
clouds from Clemens \& Barvainis (1989) and L(CDS) 1388 for the cloud with 
running number 1388 in the updated Lynds' catalogue of dark nebulae at CDS) 
and Galactic coordinates ($l $ and $b$).\\

\begin{table}
\caption[]{List of additional high-latitude clouds}
\begin{flushleft}
\begin{tabular}{lll}
\hline
Cloud Identification & $l(\degr)$ & $b(\degr)$ \\
\hline
DCld 004.9-24.6 &   4.9 &    -24.6\\
DCld 008.4-47.1 &   8.4 &    -47.1\\
DCld 009.0-46.5 &   9.0 &    -46.5\\
DCld 010.4-46.6 &   10.4&    -46.6\\
DCld 010.5-47.1 &   10.5&    -47.1\\
CB 239          &   120.61&  +24.62\\
CB 62           &   120.70&  +37.58\\
L1320           &   126.65&  +24.33\\ 
CB 61           &   127.27&  +36.96\\
L(CDS) 1388     &   147.49&  +50.14\\
DCld 313.1-28.7 &   313.1&   -28.7\\
DCld 315.1-29.0 &   315.1&   -29.0\\
DCld 315.8-27.5 &   315.8&   -27.5\\
DCld 315.8-27.5 &   315.8&   -27.5\\
DCld 317.6-28.7 &   317.6&   -28.7\\
DCld 337.5-35.3 &   337.5&   -35.3\\
DCld 337.7-35.4 &   337.7&   -35.4\\
DCld 337.9-35.5 &   337.9&   -35.5\\
DCld 338.0-26.9 &   338.0&   -26.9\\
DCld 348.6-50.6 &   348.6&   -50.6\\
DCld 348.9-46.2 &   348.9&   -46.2\\

\hline
\end{tabular}
\end{flushleft}
\end{table}

\begin{figure*}
\epsfig{figure=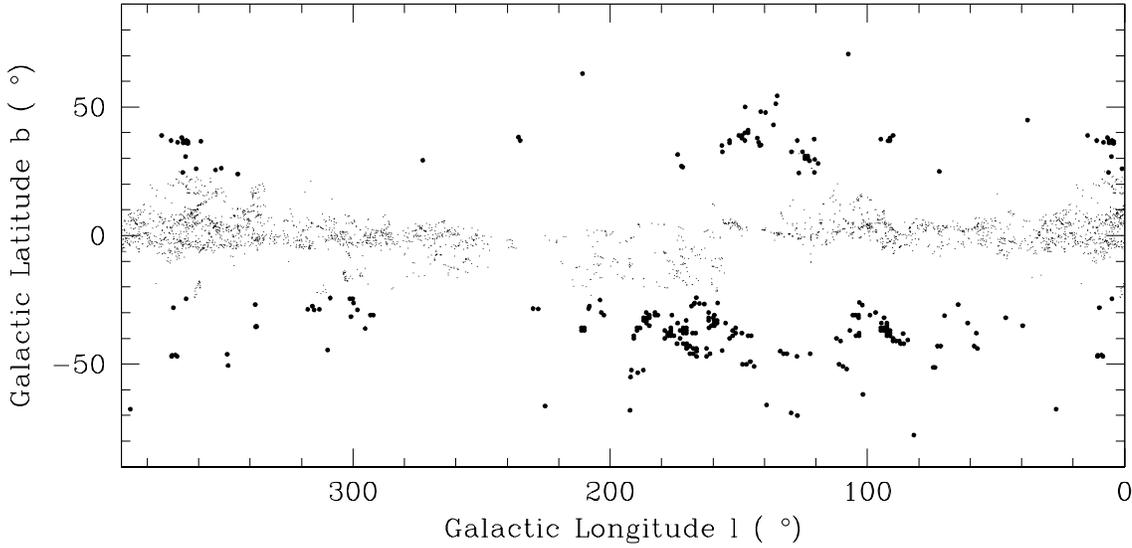,width=15cm,angle=0}
\caption{The Galactic distribution of molecular clouds: The high-latitude
clouds are shown by the larger filled symbols, while small dots represent
clouds at lower latitudes} 
\end{figure*}

\begin{figure}
\epsfig{figure=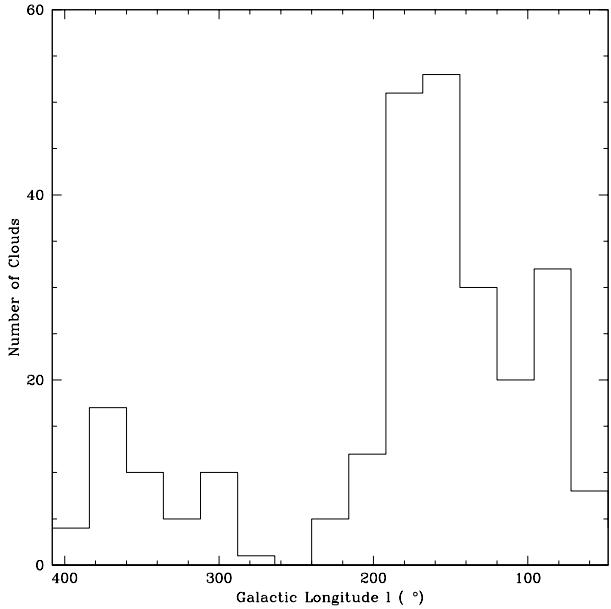,width=8cm,angle=0}
%\resizebox{\hsize}{8cm}{\includegraphics{fig2.eps}}
\caption{Histogram shows the distribution of Galactic longitudes of the
high-latitude clouds}
\end{figure}

\begin{figure*}
\epsfig{figure=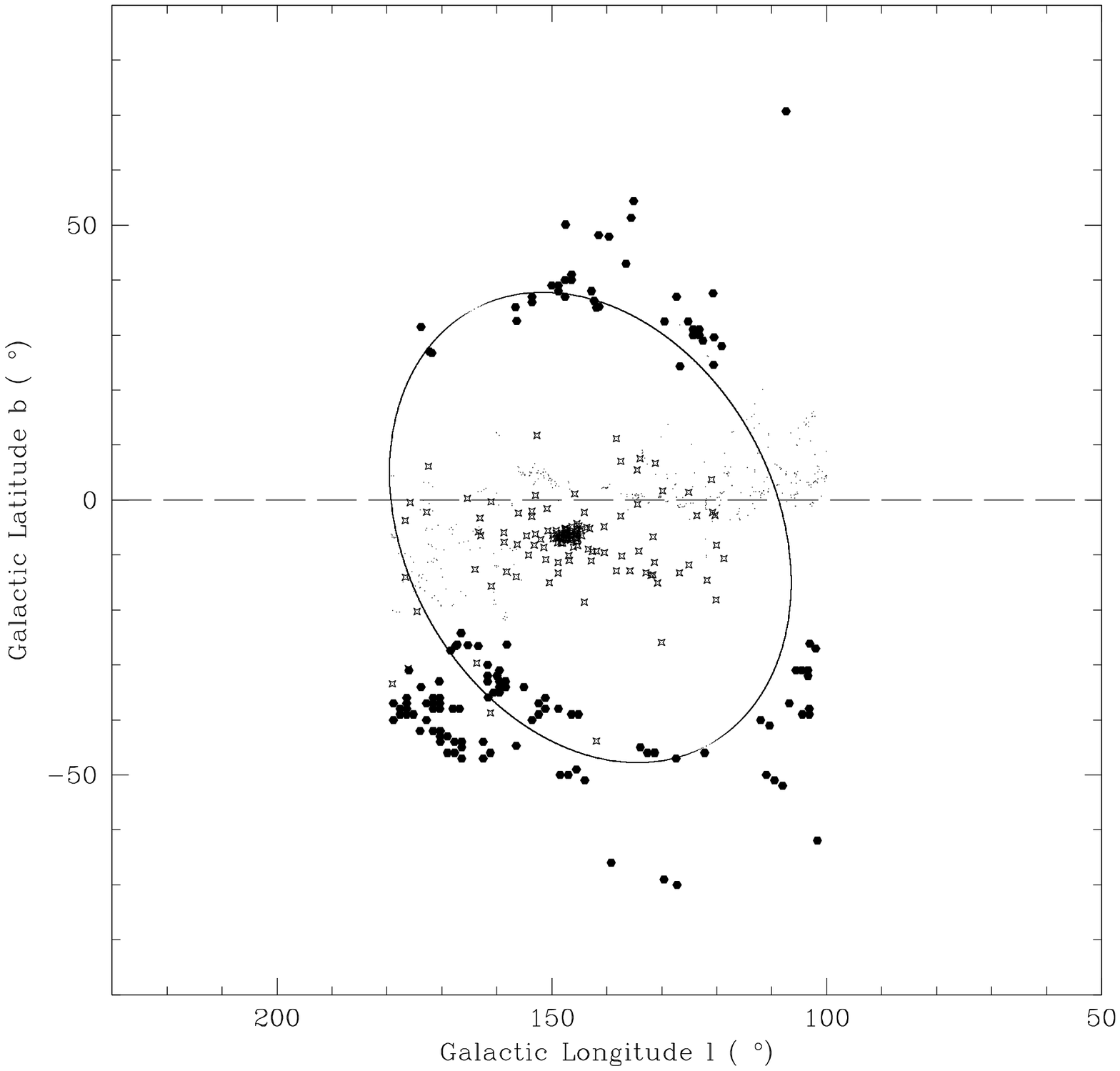,width=15cm,angle=0}
%/{\includegraphics{fig3.eps}}
\caption{The shell of high-latitude molecular clouds around the Per
OB3/Cas-Tau OB association: The OB stars of the association are shown by
open star symbols, lower latitude clouds (mostly belonging to the Gould's
Belt) by small dots, and the larger filled symbols represent the high-latitude
clouds. An ellipse is schematically fit to the shell-like distribution of
the high-latitude clouds and the Galactic equator is drawn as the
horizontal line}
\end{figure*}

Counting all the CO detections along distinct lines of sight in the NGH
and SGH surveys by HMT and MHHST as high-latitude molecular clouds
together with MHS clouds not observed or missed by these surveys, and
adding the clouds listed in Table 1, we have a total of 260 high-latitude
molecular clouds. The Galactic positions of the high-latitude clouds are
plotted in Fig. 1. Also shown in Fig. 1 are clouds at lower latitudes 
($|b| < 25^{\degr}$) from the catalogues of Lynds (1962) and Hartley et al. 
(1986). The Galactic longitude distribution of the high-latitude clouds is 
rather non uniform. There is a near-total absence of high-latitude
clouds in the longitude range $l \sim 240^{\degr} - 290^{\degr} $. As
pointed out by MHS this may represent a high-latitude analogue of the 
lower-latitude Local Bubble "tunnel" in the direction of Canis Major 
(Paresce 1984; Welsh et al. 1994). Also apparent in Fig. 1 are some clusterings
of the high-latitude clouds in both northern and the southern Galactic 
hemispheres. Fig. 2 gives a histogram of the Galactic longitude distribution
of the high-latitude clouds. The histogram shows two broad peaks in the
distribution. One in the longitude range $l \sim 70^{\degr} - 200^{\degr} $ and
the other in the range $ l \sim  290^{\degr} - 20^{\degr}$. The region with $l 
\sim 20^{\degr} - 70^{\degr} $ is relatively poor in the population of high-latitude
clouds. \\

The most prominent feature in the distribution of high-latitude clouds is found
in the longitude interval $l \sim 100^{\degr} - 180^{\degr} $. The broader 
peak in the interval $l \sim 70^{\degr} - 200^{\degr} $ contains within it a 
much sharper peak in the range $l \sim 100^{\degr} - 180^{\degr} $. This peak, 
centred at $l \simeq 150^{\degr} $, consists of 132 high-latitude clouds out of
a total of 260 clouds. The second peak in the histogram corresponding to the 
clustering of high-latitude clouds in the range $l \sim 290^{\degr} - 20^{\degr}$ is 
smaller. But this region of the sky has not been surveyed for high-latitude CO 
emission as well as the other longitude ranges accessible from observatories 
in the northern hemisphere. Clouds making up the $l \sim 100^{\degr} -
180^{\degr}$ peak in Fig. 2 seem to be distributed in a shell-like structure 
with an inner cavity that is elliptical in shape and is devoid of clouds even at
lower latitudes excepting the dark clouds belonging to the Galactic plane
and the Gould's Belt. This is shown schematically in Fig. 3. The positional 
centre of the elliptical cavity is at $l = 143^{\degr}$, $b = -5^{\degr}$. 
The projected dimensions of the cavity (major axis $\times$ minor axis)
are $\sim 90^{\degr} \times 68^{\degr}$. The major axis of the ellipse
is inclined with the Galactic plane by a large angle ($118^{\degr}$) and
is nearly perpendicular to the plane of the Gould's Belt in this region.\\

{\bf 3. OB associations and the high-latitude clouds}\\

If the Local bubble is produced by the action of supernovae and stellar winds 
from massive stars and the high-latitude clouds are formed as condensations in 
the swept-up interstellar gas (e.g. Breitschwerdt et al. 1996; Elmegreen 1988),
then it would be natural to look for any OB associations in the region of the 
system of high-latitude clouds identified above. In the longitude range $l \sim
100^{\degr} - 180^{\degr}$ and at a distance comparable to the distance 
estimates ($\sim 100 - 200$ {\rm pc}) for the high-latitude clouds, the only 
OB association is the large Cassiopeia-Taurus association whose nucleus is the 
more compact $\alpha$ Per (Per OB3) association. The $Hipparcos$ measurements 
give a distance of $177 \pm 4$ {\rm pc} for the Per OB3 association and, with 
its $3^{\degr}\times3^{\degr}$ nucleus and a halo of $\sim 10^{\degr}$, is 
centred at $l = 147^{\degr}, b = -7^{\degr}$ (De Zeeuw et al. 1999). This is 
very nearly coincident with the positional centre ($l = 143^{\degr}, 
b = -5^{\degr}$) of the elliptical cavity devoid of high-latitude clouds in 
this region. The much larger Cas-Tau association surrounds the Per OB3 
association and its members have distances ranging between $\sim 130 $ {\rm pc}
and $300$ {\rm pc} (De Zeeuw et al. 1999). We suggest here that the elliptical 
cavity and the surrounding high-latitude clouds are physically related to the 
Per OB3/Cas-Tau association. Fig. 3 illustrates this relationship, where the 
Galactic positions of the high-latitude clouds in the $ l = 100^{\degr} 
- 180^{\degr}$ range are plotted together with those of the confirmed members 
of the Per OB3/Cas-Tau association from De Zeeuw et al. (1999). The elliptical 
cavity within the inner boundary of the shell-like distribution of the clouds 
is also shown. If the high-latitude clouds in the region considered above are 
indeed related to the Per OB3/Cas-Tau association, then the mean distance to 
these clouds can be taken to be the same ($\sim 180 $ {\rm pc}) as that of the
OB association, and the linear dimensions of the shell-like structure formed
by this group of high-latitude clouds (hereafter Per OB3 shell) are $ \sim$ 280
{\rm pc} $\times$ 200 {\rm pc}. The mean shell radius is $\sim $120 {\rm pc}.\\

{\it 3.1. Kinematics of the clouds in Per OB3 shell}\\

The mean radial velocity, with respect to the local standard of rest, of the 
clouds (from radial velocity data in MHS, HMT and MHHST) is -2.15 {\rm km 
s$^{-1}$ and the velocity dispersion is  6.4 {\rm km s$^{-1}$; if the two 
intermediate velocity clouds near $l = 135^{\degr}, b = 53^{\degr}$ with 
velocities $\sim  - 45$ {\rm km s$^{-1}$ are excluded. For the Per OB3 member
stars (De Zeeuw et al. 1999) the mean radial velocity is -0.7 {\rm km s$^{-1}$
and the velocity dispersion is $\sim  4.8$ {\rm km s$^{-1}$. While a majority
of the high-latitude clouds in the range $l = 100^{\degr} - 180^{\degr} $ are 
likely members of the Per OB3 shell, a few clouds (like the intermediate 
velocity clouds mentioned above) may be unrelated to this group. Some of the 
clouds with $l \sim 170^{\degr} \pm 5^{\degr}$, $b \sim -40^{\degr} $ to $
-25^{\degr}$, having radial velocities $\sim +8$ {\rm km s$^{-1}$, similar to 
that of the near-by ($l \sim 170^{\degr}, b \sim -19^{\degr}$) lower latitude 
Taurus clouds (Taylor et al. 1987), may possibly be related to the
Taurus cloud complex. Also, a few high-latitude clouds (especially in the
southern Galactic hemisphere) outside of the longitude range $l = 100^{\degr} -
180^{\degr} $ and some clouds at lower latitudes  may actually be related to the
Per OB3 shell. However, in the present discussion we consider only 
high-latitude clouds within the range $l = 100^{\degr} - 180^{\degr}$ as 
possible Per OB3 shell members. Fig. 4 shows a plot of the radial velocities of
the clouds against the Galactic longitude. The intermediate velocity clouds 
have been excluded from this plot. Also plotted in Fig. 4 are the radial
velocities expected for the clouds at their positions ($l, b $) due to
differential Galactic rotation following the relation\\

\begin{figure}
\epsfig{figure=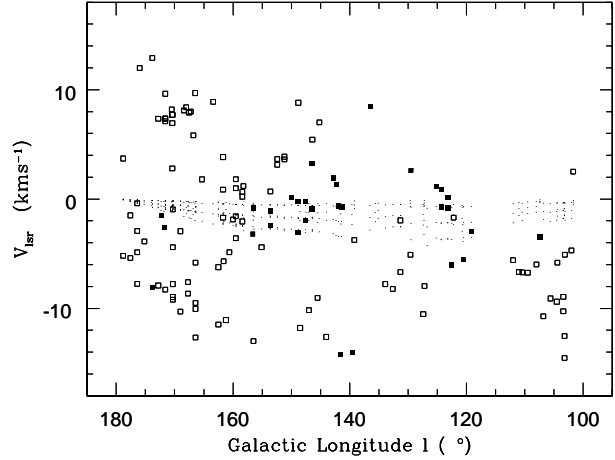,width=8cm,angle=0}
%/\resizebox{\hsize}{8cm}{\includegraphics{fig4.eps}}
\caption{Plot showing the radial velocity $V_{\rm lsr}$ as a function of
Galactic longitude for the Per OB3/Cas-Tau shell of high-latitude clouds.
Expected radial velocities due only to the differential Galactic rotation
are shown by dots for three values ( 60, 180 and 300 {\rm pc}) of distances to
the clouds. Filled squares denote clouds with $b>0$, while open squares
represent clouds with $b<0$ }
 \end{figure}

$V_{\rm lsr}(l) = A.d.sin(2l)cos^{2}(b)$\\

where $A = 17.7 $ {\rm km s$^{-1}$kpc$^{-1}$} is the Oort's constant for
$R_{\sun} = 8.5$ {\rm kpc} (Clemens 1985) and $d$ is the distance to the cloud.
For the estimated mean distance (180 {\rm pc}) and linear dimensions (mean 
shell radius $\sim$ 120 {\rm pc}) for the Per OB3 shell of clouds we have
plotted the radial velocities due to differential rotation for three values 
(60, 180 and 300 {\rm pc}) of distance $d$. Measured cloud velocities that fall
within or close to the range given by the three sets of points in Fig. 4 can
perhaps be ascribed entirely to Galactic differential rotation. It can be
seen from Fig. 4 that a large number of clouds have radial velocities
much in excess of those expected due to differential Galactic rotation.
The large observed velocity dispersion ($\sim 6.4 $ {\rm km s$^{-1}$})
must be due to large random velocities or systematic motion of the clouds. 
Some evidence for systematic motions is apparent from Fig. 4. Clouds in the 
southern Galactic hemisphere have velocities that are on an average lower 
(more negative) than those of northern clouds. This difference becomes more
pronounced if the southern clouds with $l \sim 165^{\degr} -175^{\degr} $
and velocities ($\sim +8 $ {\rm km s$^{-1}$) similar to the Taurus clouds are 
indeed related to the Taurus cloud complex and are then not members of the 
Per OB3 shell.\\  

The observed velocity distribution and the shell-like structure formed
by the Per OB3 shell of high-latitude clouds can be understood if these
clouds formed from an expanding shell of swept-up gas around the Per OB3
association. For an expansion speed of the order of $\sqrt 3 $ times the
one-dimensional radial velocity dispersion (6.4 {\rm km s$^{-1}$}), i.e. $\sim
11 $ {\rm km s$^{-1}$, and a mean distance ($\sim$ 120 {\rm pc})  of the
clouds from the centre of the Per OB3 association, the kinematic age of
the shell can be estimated to be $\sim$ 11 {\rm Myr}. Expansion from a common
centre is also supported by Fig. 5 where the observed radial velocities of
the clouds are plotted against $(1 - sin^{2}\theta/sin^{2}\theta_{\rm 
max})^{1/2}$ , $\theta$ being the angular distance of the cloud from the centre
of expansion ($l = 147^{\degr}, b = -7^{\degr}$). $\theta_{\rm max}$ ($=
70^{\degr}$) is the maximum angular separation. If the clouds are distributed 
in a shell (thin) and expansion velocity is $V_{\rm exp}$, then the observed 
radial velocity would be given by\\

\begin{figure}
\epsfig{figure=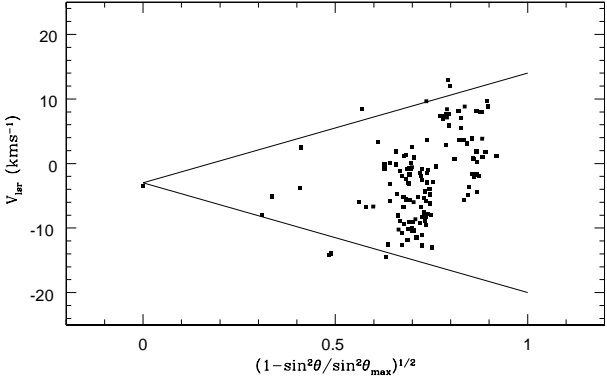,width=8cm,angle=0}
%/\resizebox{\hsize}{8cm}{\includegraphics{fig5.eps}}
\caption{Expansion of the Per OB3/Cas-Tau shell of high-latitude clouds:
The radial velocity $V_{\rm lsr}$ is plotted against $(1 -
sin^{2}\theta/sin^{2}\theta_{\rm max})^{1/2}$}
\end{figure}

{\bf $V_{\rm lsr}$ = $V_{0} \pm V_{\rm exp}\times(1 - sin^{2}\theta/sin^{2}
\theta_{\rm max})^{1/2}$}\\\\

where $V_{0}$ is the systemic velocity of the whole group (see Rajagopal
1997). For a thin-shell distribution the points in the plot would lie
along two straight lines as shown in Fig. 5. If the clouds are distributed
in a thick shell volume, then the points would lie within the envelope
defined by the two lines. It can be seen from Figs. 3 and 5 that most of the
clouds lie at projected angular distances $\sim 40^{\degr} \pm 10^{\degr}$ from
the centre and have radial velocities that are more or less uniformly
distributed between the two lines, while there is a relative paucity of 
clouds with larger angular distances having velocities that fall
between the two lines. This may indicate the presence of a thick shell of
clouds at $\sim 40^{\degr} \pm 10^{\degr}$ from the centre surrounded by a
thin shell of much larger angular radius ($\sim 70^{\degr}$). The fit
shown in Fig. 5 (with $95 \%$ of the points contained within the envelope 
defined by the two lines, and a single isolated cloud at $l = 107^{\degr}$, 
$b = 71^{\degr}$ in Fig. 3 not considered for the fit in Fig. 5) for the Per OB3
clouds gives: $V_{\rm exp} = 17 $  {\rm km s$^{-1}$} and $V_{0} = - 3 $
{\rm km s$^{-1}$}. For an outer shell radius of $\sim 200 $ {\rm pc} 
corresponding to $\theta_{\rm max} = 70^{\degr}$ used for the fit shown in 
Fig. 5, this implies an expansion age $\sim 12 $ {\rm Myr}. We will adopt an 
expansion velocity of 15 {\rm km s$^{-1}$} and an expansion age of 10 {\rm Myr}
in the following discussion. \\

It would be interesting to look for other signatures of expanding gas in
the Per OB3 shell. From the kinematics of the high-latitude clouds in this
region we have estimated an expansion velocity  $\sim  15 $ {\rm km s$^{-1}$}.
At lower latitudes (say $|b| \leq 15^{\degr}$) evidence for expansion may be 
present in the Galactic HI surveys (eg. Weaver and Williams 1973, 1974; or the 
Leiden/Dwingeloo survey: Hartmann \& Burton 1997). An expanding shell centred 
near $l = 147^{\degr}, b = -7^{\degr}$ would appear as a small disk of HI 
emission in measurements made at velocities close to the expansion velocity 
$\sim 15 $ {\rm km s$^{-1}$}, the approaching polar cap at negative velocities 
and the receding polar cap at positive velocities. Unfortunately, in the 
Galactic longitude range of the Per OB3 shell the HI gas of the general ISM 
along the line of sight is approaching us and dominates emission at negative 
velocities. At positive velocities near $\sim 15 $ {\rm km s$^{-1}$} some 
emission is seen in the Leiden/Dwingeloo survey, but this could also be due 
partly to HI gas along the line of sight having large velocity dispersion.
No clear pronounced disk of emission is evident. It is also possible that
the receding part of the shell is much weaker because there might have
been too little gas on that side of the Per OB3 association. At larger
angular distances (and higher Galactic latitudes) from the centre HI
emission would be enhanced in a shell at velocities near zero. Possible
existence of such emission may be seen in the Leiden/Dwingeloo HI maps
at velocities $\sim 0 \pm 2 $ {\rm km s$^{-1}$}, although the picture is rather
complicated with several HI arcs and filaments superimposed in projection
in this region.\\

{\it 3.2. Relationship of Per OB3 shell with the Gould's Belt}\\

The Per OB3/Cas-Tau association is also at the centre of the Gould's
Belt (Gould 1874), a flat system of young stars and interstellar matter 
within $\sim 500$ {\rm pc} of the Sun that is tilted by $\sim 18^{\degr}$ to
the Galactic plane. At lower latitudes the interstellar gas related to the
Gould's Belt is distributed in an elliptcal ring with semiaxes $\sim 360$ 
{\rm pc}$\times$ 210 {\rm pc}, while the stellar component is more extended 
with semiaxes $\sim 1000$  {\rm pc}$\times$ 700 {\rm pc}, and the system is
expanding with an expansion age estimated to be $\sim$ 35 {\rm Myr} (e.g. 
Olano , 1982; Poppel 1997 and references therein). The Per OB3 shell of 
high-latitude clouds under consideration here has smaller dimensions (semiaxes 
$\sim 140$ {\rm pc}$\times$  100 {\rm pc}) and seems to be nearly at right 
angles to the plane of the Gould's Belt. Thus, geometrically, the relatively 
compact Per OB3 association is surrounded by the more dispersed Cas-Tau 
association, which in turn is surrounded by the shell of high-latitude clouds. 
These structures, at lower latitudes, reside in the inner cavity of the Gould' 
Belt that is largely free of diffuse gas and interstellar clouds (e.g. Olano 
1982; Ramesh 1994) in an elliptical region with semiaxes $\sim 360$ {\rm pc}
$\times$ 210 {\rm pc}.\\ 

The kinematic age of $\sim$ 10 {\rm Myr} estimated here for the Per OB3
shell of high-latitude clouds is to be compared with the kinematic age of 
$\sim $ 35 {\rm Myr} for the expanding system of young stars and interstellar 
matter associated with the Gould's Belt and the age of $\sim$ 50 {\rm Myr} for 
Per OB3 association (Meynet et al. 1993). The shell of high-latitude clouds 
around Per OB3 is therefore a much younger feature than the ring of expanding 
interstellar matter associated with the Gould's Belt and the Per OB3/Cas-Tau 
association. Stellar winds and supernovae from the massive stars of
the Cas-Tau association centred around $\alpha $ Per (Per OB3), $\sim$
35 {\rm Myr} ago, could have produced the expanding ring of the Gould's Belt
(Blaauw 1956; Olano 1982). This event would also have produced shells
of gas at high galactic latitudes as two caps that oscillate perpendicular
to the galactic plane under the influence of gravity due to matter in the
galactic disc (Olano 1982). Half period for the vertical oscillation
depends on the mass density in the disc, and is estimated to be $\sim 33
\pm 3 $ {\rm Myr} (e.g. Rampino \& Stothers 1984). By now, the material of
the two caps would have therefore returned to the Galactic midplane. Also, 
this material would follow trajectories like in a fountain and is unlikely to
show a shell-like pattern (in the Galactic latitude distribution) centred
around its point of origin. The Per OB3 shell of high-latitude clouds is
perhaps produced by a more recent ($\sim $10 {\rm Myr} ago) supernova event in
the Per OB3 association which swept out the backfalling material of the
high-latitude gas-caps belonging to the Gould's Belt expansion event of
$\sim$ 35 {\rm Myr} ago. It is to be noted that the Per OB3/Cas-Tau
association still has several stars earlier than spectral type B3 ( De Zeeuw
et al. 1997) that can explode as supernovae. \\

\begin{figure}
\epsfig{figure=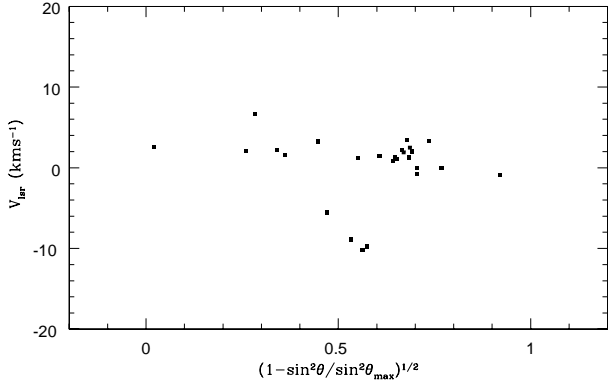,width=8cm,angle=0}
%/\resizebox{\hsize}{8cm}{\includegraphics{fig6.eps}}
\caption{The radial velocity $V_{\rm lsr}$ is plotted against $(1 - sin^{2}
\theta/sin^{2}\theta_{\rm max})^{1/2}$ for clouds associated with Sco OB2}
\end{figure}

{\it 3.3. High-latitude clouds associated with other OB associations}\\

The group of high-latitude clouds clustering around $l \sim 290^{\degr} - 20^{\degr}$ may
similarly be associated with the Sco-Cen OB association (Sco OB2) and
originate in the gas shell swept up by stellar winds and supernovae taking 
place in this young association having sub-groups with ages in the range 
$\sim$ 5 - 15 {\rm Myr} (Blauuw 1991). Unlike Per OB3 association, that has a
single centre and a clear shell-like distribution of high-latitude clouds,
Sco OB2 has several sub-groups of OB stars (Sco OB2-2: $l\sim 352^{\degr},
b\sim 20^{\degr}$, distance $\sim 145$ {\rm pc}; Sco OB2-3: $l\sim 328^{\degr},
b\sim 12^{\degr}$, distance $\sim 140$ {\rm pc}; Sco OB2-4: $l\sim 300^{\degr},
b\sim 3^{\degr}$, distance $\sim 118$ {\rm pc}; De Zeeuw et. al. 1999) with 
different ages and the shell of the high-latitude clouds is rather ill-defined.
However, the system of high-latitude clouds can be seen to envelope the 
extended OB association.\\

The system of high-latitude clouds around the Sco OB2 association may also
be expected to be expanding. Only 26 of the 44 high-latitude clouds in
this poorly surveyed region have radial velocity measurements (MHS, HMT,
MHHST). The mean radial velocity is +0.2 {\rm km s$^{-1}$} and the velocity
dispersion is 4.2 {\rm km s$^{-1}$}, which is much larger than that can be
expected to arise due to differential Galactic rotation. Fig. 6 shows a
plot of radial velocities against $(1 - sin^{2}\theta/sin^2\theta_{\rm 
max})^{1/2}$ for the Sco OB2 group of high-latitude clouds, $\theta$ being the 
angular distance of the cloud from $l = 328^{\degr}$, $b = 12^{\degr}$
representing roughly the centre of this rather extended OB association
with several subgroups. In Fig. 6, $\theta_{\rm max}$ = $58^{\degr}$.
Unlike the case of the Per OB3 clouds (Fig. 5), no clear evidence for expansion
from a common centre is apparent from Fig. 6 for the Sco OB2 clouds. More
complete CO line surveys of this region to search for high-latitude
molecular clouds and measure their radial velocities would be required to
investigate the kinematics of clouds associated with Sco OB2.\\ 

\begin{figure*}
\epsfig{figure=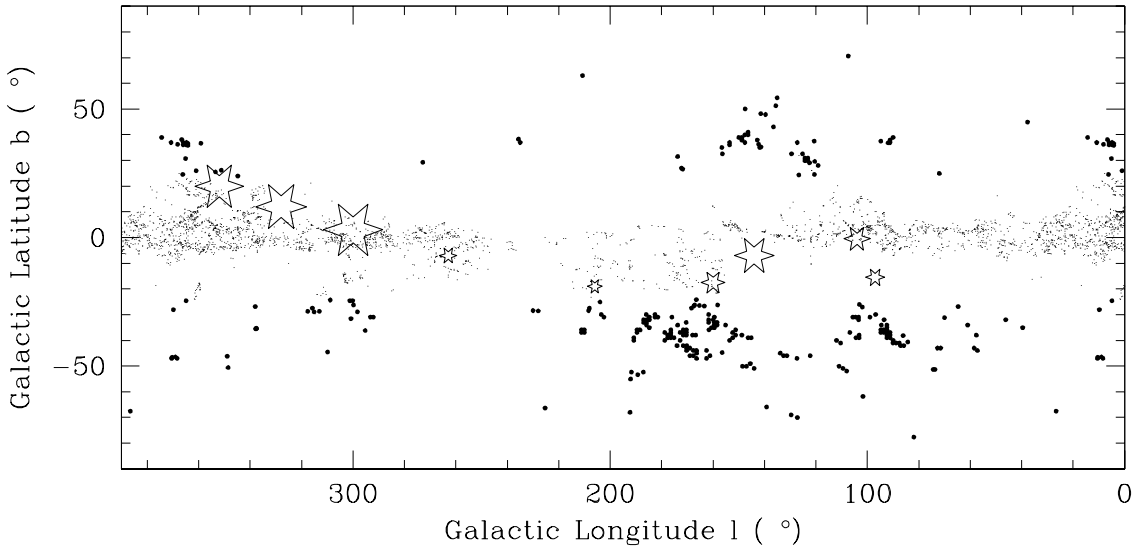,width=15cm,angle=0}
%/{\includegraphics{fig6.eps}}
\caption{Prominent OB associations within $\sim 500$ {\rm pc} are shown
together with the Galactic distribution of molecular clouds of Fig. 1. The 
OB associations are shown by the large open star shaped symbols whose
sizes are in inverse proportion to their distances}
\end{figure*}

A number of high-latitude clouds are also found in the southern Galactic
hemisphere in the longitude range $l \sim 180^{\degr} - 220^{\degr}$. It is to 
be noted that some members of the much dispersed Cas-Tau association do extend
eastward into this region and may be responsible for the formation of the
high-latitude clouds there. However, some of these clouds (especially
those at relatively lower latitudes) may be related to the Orion and
Taurus cloud complexes. The Galactic dark clouds are generally confined to
the Galactic plane within $\sim 10^{\degr}$. In some directions, however, 
clouds, other than the local high-latitude clouds being discussed here,
can be seen at relatively larger (up to $|b| \sim 20^{\degr}$) latitudes,
often making shell-like structures. These clouds could be physically
similar to the local high-latitude clouds, but much farther away. An
example is the expanding system of gas and clouds in the Gum-Vela complex
(Galactic longitudes in the range $l \sim 250^{\degr} -280^{\degr}$) centred
around ($l = 263^{\degr}, b = - 6^{\degr}$) the Vela OB2 association at a distance of
$\sim 410$ {\rm pc} (e.g. Sahu 1992; Sridharan 1992; Rajagopal 1997; 
De Zeeuw et al. 1999). This cloud complex has linear dimensions $\sim$ 210
{\rm pc} and is expanding with a speed $\sim 10$ {\rm km s$^{-1}$},
similar to the Per OB3 shell of high-latitude clouds. Were the Vela OB2 
association as close to us as the Per OB3 association (180 {\rm pc}), its 
clouds would be at galactic latitudes as large as  $-37^{\degr}$ compared to 
the observed $-16.4^{\degr}$.\\

Fig. 7 shows the galactic positions of prominent OB associations
within $\sim 500 $ {\rm pc} of the sun (taken from De Zeeuw et al. 1999)
together with the clouds of Fig. 1. As disussed earlier the clustering
of high-latitude clouds around Per OB3 and Sco OB2 is clear. At
relatively lower latitudes also the positional coincidence between OB
associations and groups of clouds showing signs of shell-like distribution
and extensions to higher latitudes is apparent. This can be seen for the
Vela OB2 (as discussed above), Ori OB1 ($l\sim 206^{\degr}, b\sim -19^{\degr}$; 
distance $\sim 470$ {\rm pc}), Per OB2 ($l\sim 160^{\degr}, b\sim -17.5^{\degr}
$; distance $\sim 300$ {\rm pc}), Lac OB1 ($l\sim 97^{\degr}, b\sim -15.5^
{\degr}$; distance $\sim 370$ {\rm pc}) and Cep OB6 ($l\sim 104^{\degr}, 
b\sim -0.5^{\degr}$; distance $\sim 270$ {\rm pc}) associations.
Being more distant than the nearby Sco OB2 and Per OB3 associations, the 
angular sizes of the shell-like cloud distributions around these OB 
associations are smaller.\\

{\bf 4. Conclusions}\\

In this paper we have considered the Galactic distributions of the
positions of the high-latitude molecular clouds and the near-by OB
associations. The conclusions can be summarised as follows.\\

(1) The majority of the high-latitude ($|b| \ge 25^{\degr}$) clouds 
 are clustered around the two closest OB associations, Per OB3/Cas-tau and
 Sco  OB2.\\

(2) The system of high-latitude clouds around Per OB3 is distributed in
 an elliptical shell with semi axes $\sim 140$  {\rm pc} $\times$ 100
 {\rm pc}. The shell is oriented nearly perpendicular to the plane of the
 Gould's Belt. It is expanding at $\sim  15 $ {\rm km s$^{-1}$} and has a 
 kinematic age $\sim$ 10 {\rm Myr}, which is much less than the kinematic age 
 ($\sim$ 35 {\rm Myr}) for the expanding ring of gas associated with the 
 Gould's Belt. It is suggested that while the primary expanding gas ring of 
 the Gould's Belt was created by the stellar winds and supernova explosions of 
 the more massive OB stars of the Per OB3/ Cas-Tau association $\sim$ 35 
 {\rm Myr} ago, the shell of the presently observed high-latitude clouds in 
 this region was produced by a more recent supernova explosion of an early B 
 type star in Per OB3 $\sim$ 10 {\rm Myr} ago as it swept up the back-falling 
 gas from the earlier events.\\

(3) More distant OB associations have also produced similar clouds at large
 heights from the Galactic plane. However, being more distant their angular
 sizes are smaller; but they can be seen as higher latitude extensions of
 these cloud complexes closer to the Galactic plane. \\

\begin{acknowledgements}

The author would like to thank the referee for several important
clarifications and suggestions. This research has made use of the SIMBAD
data base, operated at CDS, Strasbourg, France.

\end{acknowledgements}

\end{document}